\documentclass[%
 reprint,
 amsmath,amssymb,
 aps,
 prd,
longbibliography
]{revtex4-1}
\usepackage{graphicx}
\usepackage{dcolumn}

\newcommand{\GeV}{\text{\,GeV}}

\newcommand{\be}{\begin{equation}}
\newcommand{\ee}{\end{equation}}
\newcommand{\ba}{\begin{eqnarray}}
\newcommand{\ea}{\end{eqnarray}}

\renewcommand{\l}{\left(}
\renewcommand{\r}{\right)}

\begin{document}

\preprint{INR-TH}

\title{Decaying light particles in the SHiP experiment. II. \\ 
Signal rate estimates for light neutralinos}

\author{D. Gorbunov}
 \email{gorby@ms2.inr.ac.ru}
\affiliation{Institute for Nuclear Research of the Russian Academy of Sciences,
  Moscow 117312, Russia}%
\affiliation{Moscow Institute of Physics and Technology, 
  Dolgoprudny 141700, Russia}%
\author{I. Timiryasov}
 \email{timiryasov@inr.ac.ru}
\affiliation{Institute for Nuclear Research of the Russian Academy of Sciences,
  Moscow 117312, Russia}%
\affiliation{Physics Department, Moscow State University, Vorobievy Gory,  
Moscow 119991, Russia}

\begin{abstract}
Considering the supersymmetric models with light neutralino and
$R$-parity violation, we perform estimates of the signal
rate expected at the recently proposed fixed target SHiP experiment
exploiting the CERN SPS beam of 400 GeV protons. We extend the existing
studies by introducing new production channels (in particular through
the beauty mesons) and decay modes. We also constrain the model parameter
space from analysis of negative results of the CHARM experiment. 
\end{abstract}

\maketitle

\section{Introduction\label{sec:intro}}

Supersymmetric (SUSY) extensions of the standard model of particle physics
(SM) provide taming of the radiative corrections to the Higgs
boson mass (for a review see \cite{Nilles:1983ge}). 
The crucial role is played by the superpartners of the SM
particles, whose contributions cancel the quadratically divergent
quantum corrections of the SM particles. Therefore, one naturally
expects to observe (some of) the superpartners at the TeV
mass scale. Testing this prediction is one of the main tasks of
run 2 at the LHC. 

However, there can be {\it renegades} in the SUSY world with masses
(much) below the TeV scale and sufficiently suppressed interactions
with the SM fields, so they are missed by numerous previous
searches and are beyond the reach of the LHC. The renegade hunting can be performed at a
beam-dump experiment, where the high statistics of
proton(electron)-proton collisions might compensate for smallness of
the couplings, so the new light particles can be 
produced. Recently a proposal has been submitted
\cite{Bonivento:2013jag,Anelli:2015pba} to build the new
experiment at CERN with a 400\,GeV SPS proton beam. Originally motivated
as a facility to search for sterile neutrinos (heavy neutral leptons) of ${\cal O}(1)$\,GeV
mass \cite{Gorbunov:2007ak,Gninenko:2013tk,Bonivento:2013jag}, lately it has
been recognized as a universal tool to probe various models
predicting light, sufficiently long-lived, neutral particles; it has been named
SHiP (Search for Hidden Particles) \cite{Anelli:2015pba}. The SHiP
physics case is presented in a separate paper \cite{Alekhin:2015byh},
including a number of the SUSY renegades. 

In this work, we consider the light unstable neutralino in
supersymmetric models with $R$-parity violation (for reviews, see
\cite{Barbier:2004ez,Dreiner:1997uz,Mohapatra:2015fua}).  
$R$-parity is a discrete multiplicative
symmetry ascribing factor,
\begin{equation}
R_p=(-\textbf{1})^{3B+L+2S}\,,
\end{equation}
to any particle of baryon charge $B$, lepton charge $L$, and spin $S$. 
All SM fields (including scalars of the extended Higgs sector) have
$R_p=+1$, while their superpartners have $R_p=-1$. If $R$-parity is 
conserved, superpartners may only be created in pairs.  
$R$-parity guarantees stability of the lightest supersymmetric
partner (LSP) which becomes a candidate to form the dark matter
component of the Universe. 

However, the theoretical grounds of $R$-parity have been questioned
(see, e.g., Refs \cite{Hall:1983id,Perez:2013usa}), and one can
introduce $R$-parity violating (RPV) terms in the Minimal Supersymmetric
extension of the SM (MSSM). These terms trigger decays of
superpartners into SM particles, in particular, decays of the LSP. The
latter may be the lightest mass eigenstate in the sector of neutral
fermion superpartners called neutralinos. If the neutralino is
sufficiently light, it is an example of the renegades to be searched
for at the SHiP experiment. There, the neutralinos can be produced
either directly in proton scattering off the target material or
indirectly via decays of secondary hadrons, and they later decay into SM
particles exhibiting signatures very similar to those of sterile
neutrinos \cite{Gorbunov:2007ak}. Thus, the procedures applied in data
analysis to probe both models are very similar. The difference
between the two models is expected in the production channel pattern
and in decay channel patterns, so that the momentum spectra of each
final state and the weights of the different final states
($K^\pm\mu^\mp$, $\mu^+\mu^-\nu$, etc) are generically not the same in
the two models.

Light neutralino phenomenology has previously been studied in the SHiP
physics paper\,\cite{Alekhin:2015byh}. Here, we considerably extend
this study by including more production (from neutral charm mesons and
from beauty mesons) and decay ($\pi^\pm l^\mp$, $l^\pm l^\mp\nu$) channels and by
obtaining limits on the model parameter space from analysis of the published
results of the CHARM experiment \cite{Bergsma:1983rt,Bergsma:1985is}.

The paper is organized as follows. In
Sec.\,\ref{sec:production_mechanisms} we introduce the model and
calculate the neutralino production rates and decay rates for a number
of channels. In Sec.\,\ref{sec:neutralino-signal} we give the estimate
of the number of signal events expected in the SHiP fiducial
volume. In Sec.\,\ref{sec:ship_bounds_on_rpv} we present the
sensitivity of the SHiP experiment to the model parameters (assuming
zero background) and find new limits on the model parameters from the
published results of the CHARM experiment. We summarize in
Sec.\,\ref{sec:Conclusion}.


\section{Supersymmetry with RPV: production and decays of light neutralinos}
\label{sec:production_mechanisms}

$R$-parity is explicitly violated by the following terms in the MSSM
superpotential  
\begin{align}
\label{RPVW}
W_{\not R_{p}}&=\lambda_{ijk}\epsilon_{ab}L_{i}^aL_{j}^b E_{k}^C
+\lambda ^{_{\prime }}_{ijk}\epsilon_{ab}L_{i}^aQ_{j}^b D_{k}^C\\
+&\lambda ^{_{\prime \prime }}_{ijk}\epsilon_{\alpha\beta\gamma}
U_{i}^{C\,\alpha} D_{j}^{C\,\beta} 
D_{k}^{C\,\gamma}+\mu_i \epsilon_{ab}L^a_i H^a_U\,,
\label{RPVW2}
\end{align}
where dimensionless couplings $\lambda_{ijk}$ and mass parameters
$\mu_i$ ($i,j,k$ run over the three matter generations) 
characterize violation of $R$-parity, the superscript
$C$ refers to the charge conjugated fields, indices 
$a,b=1,2$ indicate the $SU(2)_W$ doublet components ($L$ and $Q$ are
lepton and quark doublets; $E$, $D$, and $U$ are lepton , down-type, and
up-type quark singlets, respectively), while
$\alpha,\beta,\gamma$ count $SU(3)_C$ triplet components;
$\epsilon_{ab}$ and $\epsilon_{\alpha\beta\gamma}$ are fully antisymmetric
$2\times2$ and $3\times3\times3$ tensors.   
Now, if all the terms \eqref{RPVW}, \eqref{RPVW2} are present, they initiate
the fast proton decay. This process can be forbidden with some
discrete remnant of $R$-parity. In particular, the baryon triality
\cite{Dreiner:2006xw} 
forbids the first term in \eqref{RPVW2} and hence keeps the proton stable. 
In what follows we concentrate on the phenomenology of the RPV terms
in Eq.\,\eqref{RPVW} and neglect the terms in Eq. \eqref{RPVW2}.
Whereas the lightest neutralino should be heavier than $46$\,GeV in the constrained MSSM with 
five parameters \cite{Agashe:2014kda}, 
the authors of Ref. \cite{Dreiner:2009ic} show that this bound can be relaxed and, even a
massless neutralino is possible. 
In this study we consider models with the mass of the lightest neutralino in a GeV range.

Neutralinos as the LSPs could be created in decays of heavier sparticles. 
Missing momentum carried out by the LSP, which escaped from a detector, remains
one of the main signatures in collider searches of supersymmetry. 
The proposed center-of-mass energy of the SHiP experiment 
$\sqrt{s}=27.4$\,GeV is too low to create heavy superpartners with masses
above the electroweak scale, as we anticipate  
from LEP-II, Tevatron, and LHC run I.  
One can check that the neutralino direct production in
the proton-proton scattering is negligibly low. 
Therefore, here we study an indirect production of neutralinos in 
decays of heavy mesons via $R$-odd couplings $\lambda'$ in \eqref{RPVW}.
These $R$-odd couplings also lead to neutralino decays 
into the ordinary SM particles.

\subsection{Neutralino production in decays of heavy mesons} 
\label{sub:neutralino_production_in_decays_of_heavy_mesons}

\begin{figure}[!htb]
  \centerline{
  \includegraphics[width=0.13\textwidth]{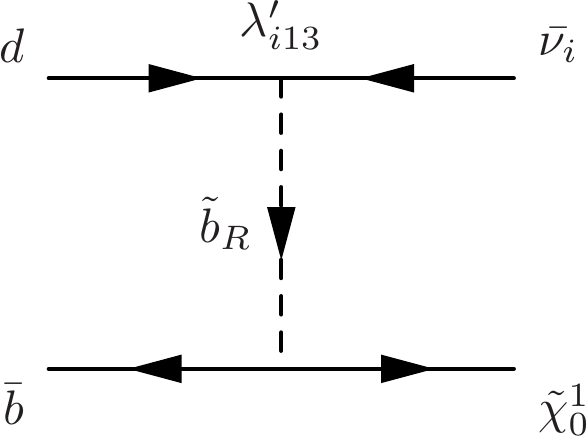}
\hskip 0.05\textwidth
 \includegraphics[width=0.13\textwidth]{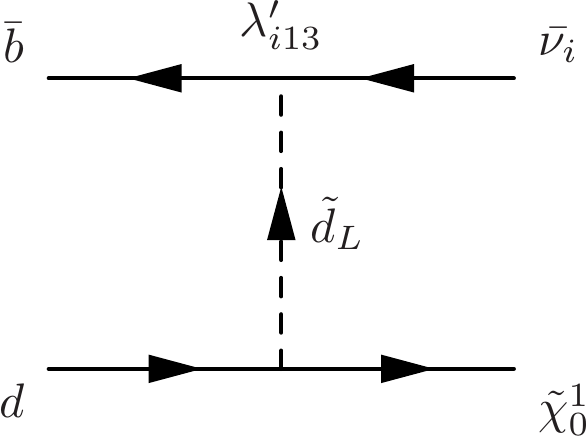}
 \hskip 0.05\textwidth
 \includegraphics[width=0.13\textwidth]{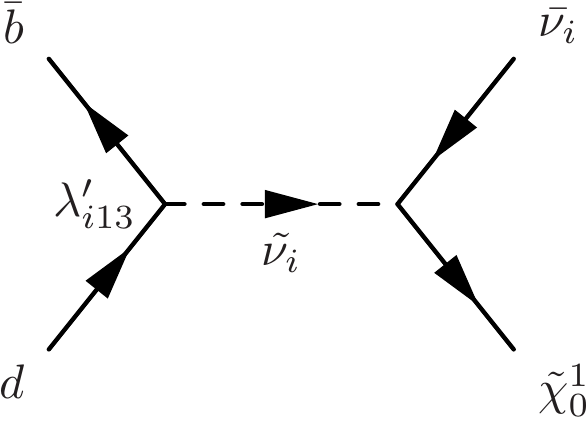}}
  \caption{Typical Feynman graphs of neutralino production in 
meson decay via $R$-parity-violating couplings $\lambda'$.}
  \label{fgraphs}
\end{figure}

Light enough neutralinos $\tilde{\chi}^0_1$  
can be produced in decays of heavy mesons (charm 
$D$ and beauty $B$) provided that we have $R$-parity-violating 
coupling $\lambda'$ as shown in Fig.\,\ref{fgraphs}.

Expressions for the partial widths of $B_0$ and $B^+$ meson decays can 
be found in Ref.\,\cite{Dedes:2001zia} (for details and derivation of
similar expressions see also Ref.\,\cite{Choudhury:1999tn}). For 
a neutralino being a pure bino state, they read 
\begin{widetext}
\begin{align}
\Gamma\l B^0_{d}\to\bar{\nu}_i\tilde{\chi}^0_1\r =\frac{\lambda'^2_{i13} g'^2 f^2_B M^2_{B_0} p_{cm}}{128\pi(m_d+m_b)^2}
\left [
  \frac{Y_{\nu_i}}{M_{\tilde{\nu}_i}^2}-\frac{Y_d}{M_{\tilde{d}_L}^2}
+\frac{Y_b^*}{M_{\tilde{b}_R}^2} \right]^2 
\l M^2_{B^0}-M^2_{\tilde{\chi}^0_1}\r,  \label{eqn:B0}\\
\Gamma\l B^+\to\ell^+_i\tilde{\chi}^0_i\r =\frac{\lambda'^2_{i13} g'^2 f^2_B M^2_{B^+} p_{cm}}{64\pi(m_u+m_b)^2}
\left [
  \frac{Y_{l_i}}{M_{\tilde{l}_i}^2}-\frac{Y_u}{M_{\tilde{u}_L}^2}+
\frac{Y_b^*}{M_{\tilde{b}_R}^2} \right]^2 
\l M^2_{B^+}-m^2_{\ell_i}-M^2_{\tilde{\chi}^0_1}\r,
\label{eqn:B+}
\end{align}
\end{widetext}
where $p_{cm}$ is the 3-momentum of outgoing particles in the rest frame 
of decaying meson, $m_u$, $m_d$, $m_b$ are quark masses, 
$m_{\ell_i}$ is mass of the final state lepton (electron or muon),
$M_{\tilde{\nu}}$,  $M_{\tilde{d}_L}$, $M_{\tilde{b}_R}$, \dots 
are sfermion masses, 
$Y_{\nu_i}$, $Y_d$, $Y_b$, \dots 
are corresponding hypercharges (coming from neutralino couplings in
the pure bino limit), $g'$ is $U(1)_Y$ gauge coupling constant, $M_{\tilde{\chi}^0_1}$ is neutralino mass, 
$M_{B^+}$, $M_{B^0}$ are $B^+$ and $B^0$ masses, respectively, and
$f_B=204\pm30$ MeV \cite{Agashe:2014kda} is the $B$-meson decay
constant. Note that definition of the constant $f_B$ in
\cite{Dedes:2001zia} 
differs from that by the Particle Data Group \cite{Agashe:2014kda}. Therefore, formulas
from \cite{Dedes:2001zia} should be multiplied by $1/2$ for our
choice of $f_B$. 

Without loss of generality, hereafter we assume the common mass scale of sfermions
$M_{\tilde{f}}\equiv
M_{\tilde{\nu}_i}=M_{\tilde{d}_L}=\dots=M_{\tilde{b}_R}$    
This assumption simplifies further phenomenological
treatment since Eqs.\,\eqref{eqn:B0} and \eqref{eqn:B+} turn into:
\begin{widetext}
\begin{align}
\Gamma\l B^0_{d}\to\bar{\nu}_i\tilde{\chi}^0_1\r &=\l \frac{\lambda'_{i13}}{M^2_{\tilde{f}}}\r^2\frac{9g'^2f^2_Bm^2_{B^0}p_{cm}}{512\pi(m_d+m_b)^2}
\l M^2_{B^0}-M^2_{\tilde{\chi}^0_1}\r,  \label{eqn:B0simp}\\
\Gamma\l B^+\to\ell^+_i\tilde{\chi}^0_i\r &=\l \frac{\lambda'_{i13}}{M^2_{\tilde{f}}}\r^2 \frac{9g'^2f^2_Bm^2_{B^+}p_{cm}}{256\pi(m_u+m_b)^2}
\l M^2_{B^+}-m^2_{\ell_i}-M^2_{\tilde{\chi}^0_1}\r
\label{eqn:B+simp}
\end{align}
\end{widetext}
and the production rates of neutralinos are proportional to the 
squared combination $\lambda'_{i13}/M^2_{\tilde{f}}$. Note that our assumption is not related to any particular
pattern of mass spectrum of RPV SUSY.
In a more general case of different masses of superpartners, the dominant contribution will come from the 
intermediate sfermion with the highest value of $\lambda'/M^2_{\tilde{f}}$, e.g., from the lightest sfermion, if all RPV
couplings are of the same order.

Expressions similar to \eqref{eqn:B0simp} and \eqref{eqn:B+simp} can be obtained for decays of $D$ mesons with obvious replacement of couplings 
$\lambda'_{i13} \to \lambda'_{i21}$ and replacement of the corresponding quark flavors.


\subsection{Decay pattern} 
\label{sec:decay_pattern}
We study three different decay channels of light neutralinos with two
charged particles in the final state: three-body leptonic decay via
$\lambda_{ijk}$, two-body semileptonic decay into a charged pion via
the coupling $\lambda'_{i11}$, and two-body semileptonic decay into a charged
kaon via $\lambda'_{i12}$. These states yield the detectable at SHiP
signature of two charged particles originated from a single vertex. 

The amplitude of the leptonic neutralino decay through a
virtual slepton or sneutrino was calculated in Ref. \cite{Dreiner:1999qz}.
The decay width in the pure bino limit (neglecting masses of the final-state
particles) reads
\begin{equation}
	\Gamma(\tilde{\chi}^0_1\to \bar{\nu}_i\ell_j^+\ell_k^-)=\l \frac{\lambda_{ijk}}{M^2_{\tilde{f}}}\r^2
	\frac{3 g'^2 M^5_{\tilde{\chi}^0_1}}{4096 \pi^3}\,.
	\label{leptonic_decay}
\end{equation} 
The rate of semileptonic decay can be calculated analogously to
Eq.\,\eqref{eqn:B0}. In the pure bino limit one has \cite{Alekhin:2015byh}
\begin{widetext}
\begin{equation}
    \Gamma(\tilde{\chi}^0_1\to K^- \ell^+_i)=\frac{9}{256 \pi} \l \frac{\lambda'_{i12}}{M^2_{\tilde{f}}}\r^2 
  \frac{g'^2 f_K^2 m^4_{K^+}\l M_{\tilde{\chi}^0_1}^2-m_{K^+}^2-{m_\ell^i}^2\r p_{cm}}
  {M_{\tilde{\chi}^0_1}^2(m_s+m_u)^2}
	\label{semileptonic_decay2}
  \end{equation}
\end{widetext}
and the expression analogous to \eqref{semileptonic_decay2} for 
$\tilde{\chi}^0_1\to \pi^- \ell^+_i$, which is proportional to
$|\lambda'_{i11}|^2$. 
One can see from Eqs.\,\eqref{leptonic_decay} and
\eqref{semileptonic_decay2} 
that the decay rates of a neutralino, as well as its production rates 
[Eqs. \eqref{eqn:B0simp} and \eqref{eqn:B+simp}], 
are proportional to the factor $\l \lambda/M^2_{\tilde{f}} \r^2$.

If $\tilde{\chi}_1^0$ is produced in decays of $D$ mesons via
$\lambda'_{i21}$ then there is one additional decay channel
$\tilde{\chi}^0_1\to K^0 \nu_i$. A special study is needed to
understand whether it can be  distinguished
from background since there are no charged particles in the final state 
\footnote{There is a possibility for a part of events where the kaon
  decays to $\pi^+\pi^-$ inside a detector fiducial volume.}
but it affects the neutralino lifetime. Hence we take into account
the decay rate 
\begin{widetext}
\begin{equation}
	\Gamma(\tilde{\chi}^0_1\to K^0 \bar{\nu}_i)= \frac{1}{512 \pi} \l \frac{\lambda'_{i21}}{M^2_{\tilde{f}}}\r^2 
	\frac{g'^2 f_K^2 m^4_{K^0}\l M_{\tilde{\chi}^0_1}^2-m_{K^0}^2\r p_{cm}}
	{M_{\tilde{\chi}^0_1}^2(m_s+m_d)^2}  \label{semileptonic_decay} 
  \end{equation}
\end{widetext}
and a similar one for 
$\tilde{\chi}^0_1\to \pi^0 \nu_i$ (which is proportional to
$|\lambda'_{i11}|^2$) in further investigations.

\section{Signal event rates}
\label{sec:neutralino-signal}

The goal of the present paper is to estimate the event rate of light
neutralino decays within the SHiP detector. We consider the light
neutralinos created in decays of heavy mesons. These heavy
mesons are produced, in turn, by 400 GeV protons scattering off the
target material.

For the production cross section of a particle (a neutralino in our case) 
with 3-momentum $\vec{p}$ created in a decay of the heavy meson $H$ ($H = D, B$) 
with 3-momentum $\vec{k}$, 
one has:
\begin{equation}
  \frac{d^3 \sigma}{d p d\theta_p d\phi_p} = \mathcal{B} \int d^3 k 
  f(\vec{p},\vec{k}) \frac{d^3\sigma_H}{d k d\theta_k d\phi_k},
  \label{chi_prod}
\end{equation}
where $p\equiv|\vec{p}|,k\equiv|\vec{k}|$, $\mathcal{B}$ is the
branching ratio of $H$ two-body decay to neutralino,
$f(\vec{p},\vec{k})$ is the momentum distribution of a neutralino, and
$\frac{d\sigma_H}{d k d\theta_k d\phi_k}$ is the differential
production cross section of meson $H$ in $pp$ collisions. All the
momenta in Eq. \eqref{chi_prod} are given in the laboratory frame.  In
the rest frame of the meson $H$ (denoted by an asterisk), neutralino
3-momentum is uniformly distributed and one finds:
\begin{equation}
   f(\vec{p}^{\,*},0)=\frac{1}{2\pi p^{\,*}}  \delta({p}^{*\mu} {p}^*_\mu-m^2),
   \label{fCM}
 \end{equation} 
where $p^{*\mu}$ is the 4-momentum of the neutralino in the rest frame of the decaying meson. 
Note that the value of 
$p^{\,*}\equiv|\vec{p}^{\,*}|$ 
is fixed by kinematics of the two-body decay. In order to boost
expression \eqref{fCM} to the laboratory frame, one should multiply it
by the appropriate Jacobian and express
$\vec{p}^{\,*}=\vec{p}^{\,*}(\vec{p}, \vec{k})$ in terms of the 3-momenta 
of the neutralino ($\vec{p}$) and the decaying meson ($\vec{k}$) 
in the laboratory frame. 

The differential cross section of D-meson production in pp interactions at the
center-of-mass energy $\sqrt{s}=27.4$\,GeV, which is relevant for the SHiP setup, 
was measured by the LEBS-EHC Collaboration
\cite{AguilarBenitez:1987rc}.  It has been found that the differential
production cross section is well represented by the empirical form
\cite{AguilarBenitez:1987rc}
\begin{equation}
  \frac{d\sigma_D}{dx_Fdp^2_T} = \frac12\l \sigma(D/\bar{D})(n+1)b \r \l1-|x_F|\r^n \exp(-bp_T^2)
  \label{Aguilar}
\end{equation}
with $n=4.9\pm 0.5$, $b=(1.0\pm 0.1)\GeV^{-2}$. We adopt the same value of the
inclusive $D/\bar{D}$ cross section $\sigma(D/\bar{D})=18 
\mu$b \cite{Alekhin:2015byh}. The differential cross section in \eqref{Aguilar} 
depends on transverse $p_T$ and longitudinal $p_L$ components of
3-momenta through $x_F=2p_L/\sqrt{s}$ and can be related to that used
in Eq.\,\eqref{chi_prod} as follows:
\begin{equation}
 \frac{d^3\sigma_D}{d k d\theta_k d\phi_k}=\frac{4 k^2 \sin\theta_k}{\sqrt{s}}\frac{d^3\sigma}{dx_F dk_T^2 d\phi_k} .
\end{equation}

The beauty production cross section has not been measured in the 
interesting energy region. 
Therefore, following the SHiP Collaboration \cite{Alekhin:2015byh} we
extrapolate existing data \cite{Lourenco:2006vw} to estimate the number of
produced $B$ mesons.  To estimate the angular distribution of produced 
$B$ mesons, we employ the theoretical results from Ref.\,\cite{Mangano:1992kq}
together with the Lund fragmentation model \cite{Andersson:1983ia}.

The probability of $\tilde{\chi}_1^0$ decay inside 
the fiducial volume of the SHiP detector is 
\begin{eqnarray}
w_{\text{det}}= e^{-l_\text{sh}/l_{\tilde{\chi}_1^0}}
(1-e^{-l_\text{fid}/l_{\tilde{\chi}_1^0}})\simeq \frac{l_\text{fid}}{l_{\tilde{\chi}_1^0}},
\label{bb}\\
l_{\tilde{\chi}_1^0}=\frac{p}{M_{\tilde{\chi}_1^0} \Gamma} \nonumber,
\end{eqnarray}
with $l_\text{sh}$ denoting the muon shielding length (the distance
between the collision point and the detector, 63.8\,m 
for SHiP \cite{Anelli:2015pba}) and
$l_\text{fid}$ referring to the length of the detector 
fiducial volume (60\,m); 
the second equality in \eqref{bb} is valid when $l_\text{sh}\ll
l_{\tilde{\chi}_1^0}$.

The proposed geometry of the SHiP detector \cite{Anelli:2015pba} is a 60\,m length
  cylindrical vacuum tank with an elliptical section of $x$ and $y$ semiaxes 2.5\,m and 5\,m long, respectively.  
  In the following estimates of
  the signal event numbers, we utilize a more conservative fiducial 
volume that is a cone formed by 
 the vertex in the target and the $5\text{~m} \times 10$ m ellipse  
at the very end of the fiducial volume.  It covers part of the
elliptical section. We argue that this choice is quite reasonable since neutralino decay products should be tracked by
the detector
placed at the end of the vacuum tank.
We select only neutralinos $\tilde{\chi}_1^0$ with 3-momenta
inside the cone region 
described above. 

The number of neutralino decays within the detector is given by
\begin{equation}
	N=\frac{N_{\rm POT}}{\sigma_{\rm pp, total}}\int_{\text{cut}} w_{\text{det}} \frac{d\sigma_{\tilde{\chi}_1^0}}{dp d\theta d\phi} d^3p
  \label{Nsig}
\end{equation}
where the distribution of neutralinos over the 3-momentum is 
defined in \eqref{chi_prod} and ``cut'' refers to the constraint on the neutralinos'
 3-momenta described above
and $N_{\rm POT} = 2\times10^{20}$ is the number of protons on target during 5 years of operation 
\cite{Alekhin:2015byh}. 

We consider the four production channels of light neutralinos
described in Sec. \ref{sec:production_mechanisms} : $D^\pm$, $D^0$
decays via coupling $\lambda'_{i 2 1}$ and $B^\pm$, $B^0$ decays via coupling 
$\lambda'_{i 3 1}$.  According to the SHiP technical proposal
\cite{Anelli:2015pba}, at least two charged particles are required
to distinguish the signal from a background. Hence we consider three decay
channels: $\tilde{\chi}^0_1\to \ell_i^+\bar{\nu}_j\ell_k^-,\quad
\tilde{\chi}^0_1\to K^- \ell^+_i,\quad \tilde{\chi}^0_1\to \pi^-
\ell^+_i$, driven by $\lambda_{ijk},\quad \lambda'_{i12},\quad
\lambda'_{i11}$ correspondingly.  As a result, we have six
combinations of couplings that can be tested by SHiP.

\begin{figure}[!htb]
    \includegraphics[width=0.4\textwidth]{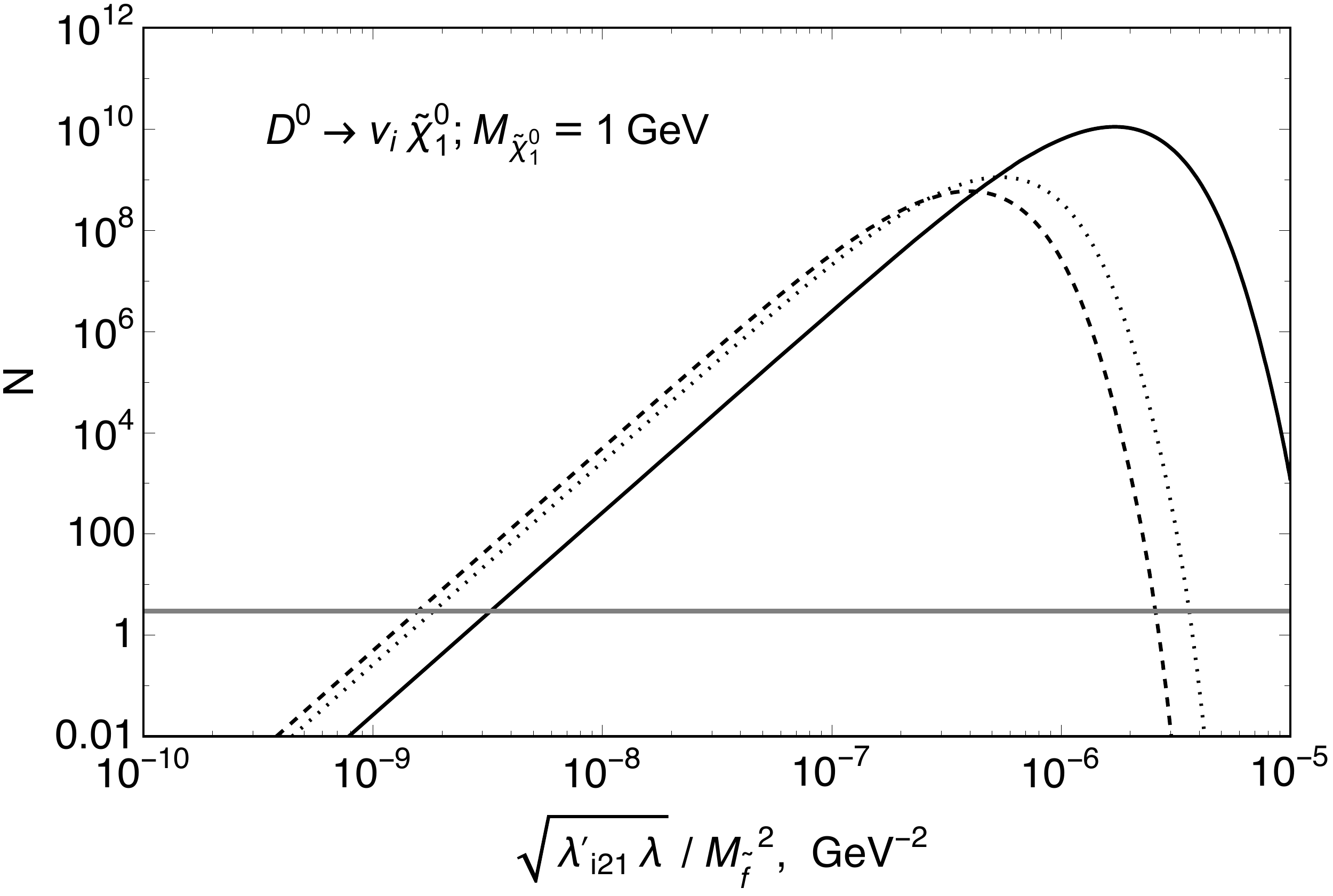}
 \vskip 0.2cm
\includegraphics[width=0.4\textwidth]{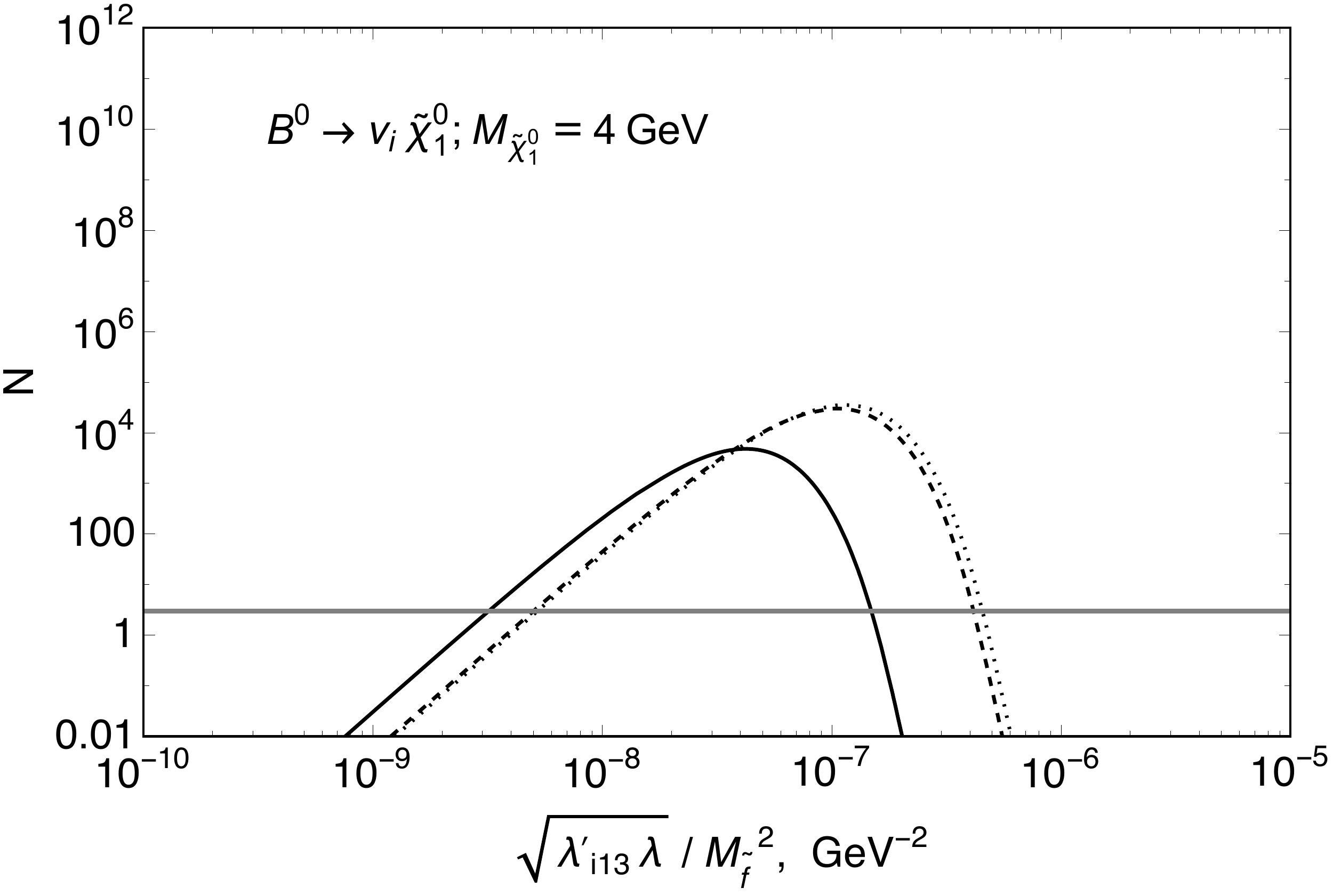}
  \caption{Number of signal events for neutralinos produced in $D^0$ decay (upper panel)
  and in $B^0$ decay (lower panel). Decays of the neutralino to 
  $e^+ \nu_i e^-$ (solid line), to $\pi^+ \ell^-$ (dashed line), and to 
  $K^+ \ell^-$ (dotted line) are considered. $\lambda$ stands for
  the appropriate RPV couplings. 
The horizontal line represents three events: their absence implies
  an upper limit on the model parameters placed at $95\%$ confidence
    level within the Poisson statistics. 
  \label{LofM}}
\end{figure}

\begin{figure}[!htb]
  \includegraphics[width=0.43\textwidth]{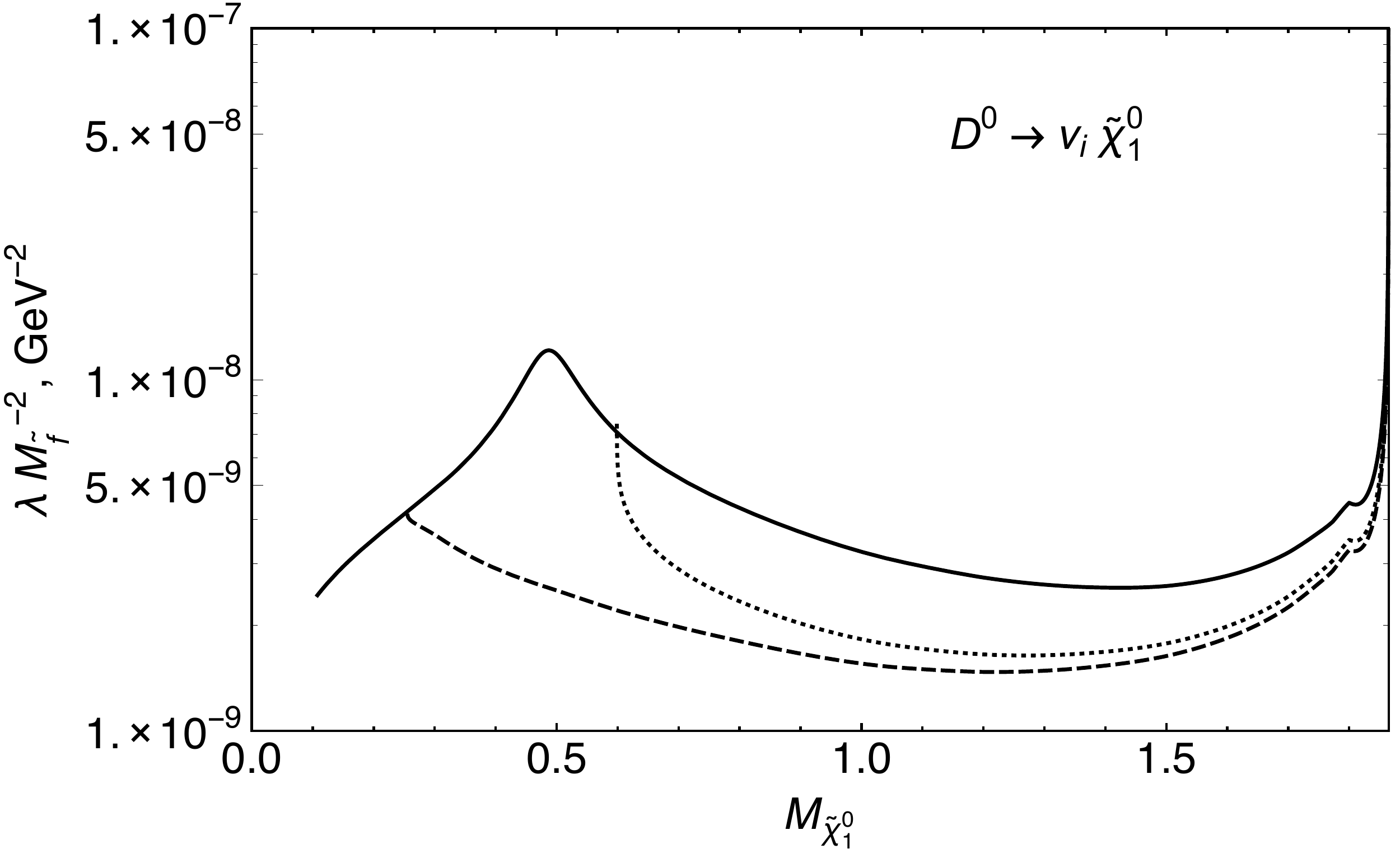}
   \vskip 0.2cm
  \includegraphics[width=0.43\textwidth]{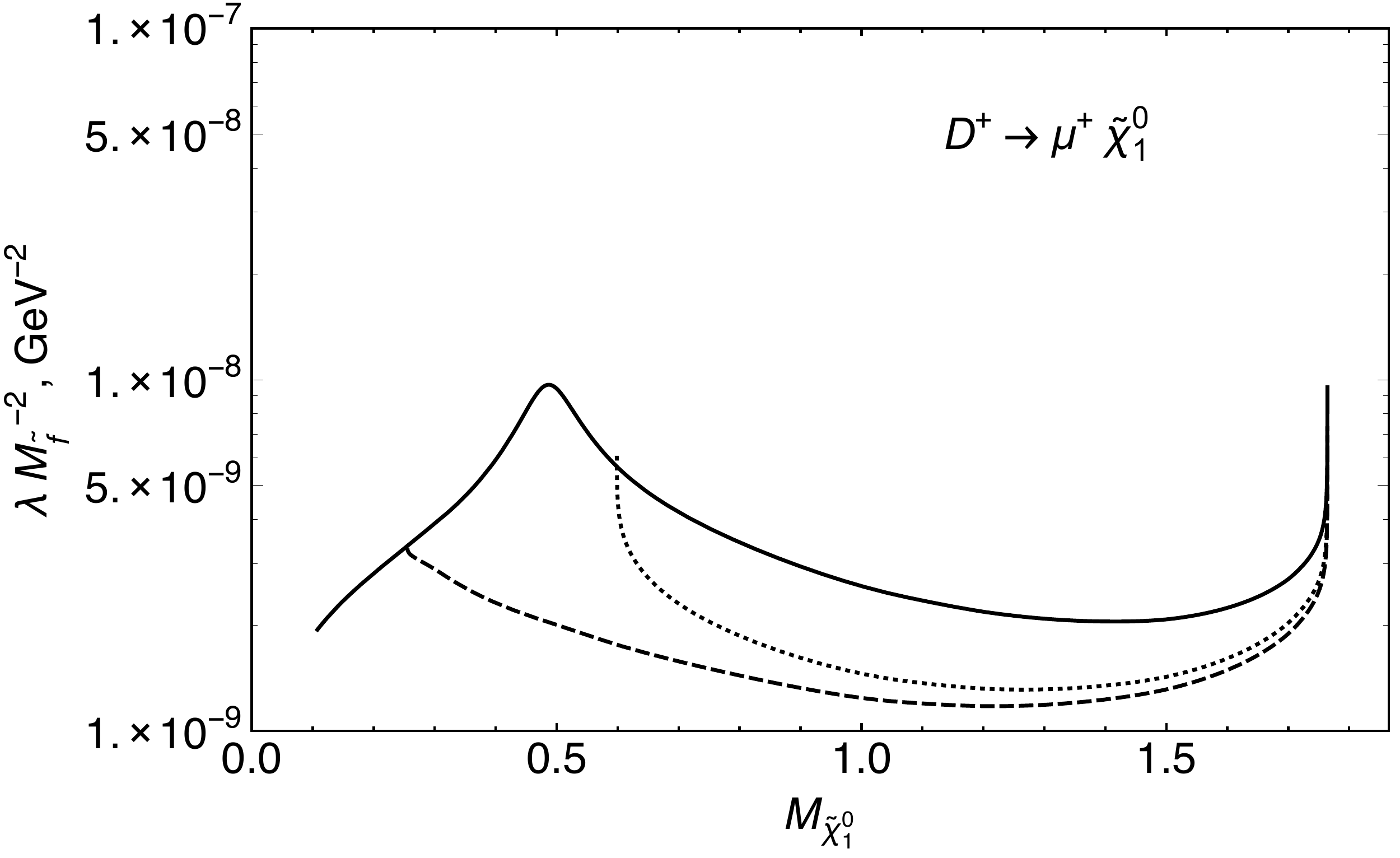}
   \vskip 0.2cm
\includegraphics[width=0.43\textwidth]{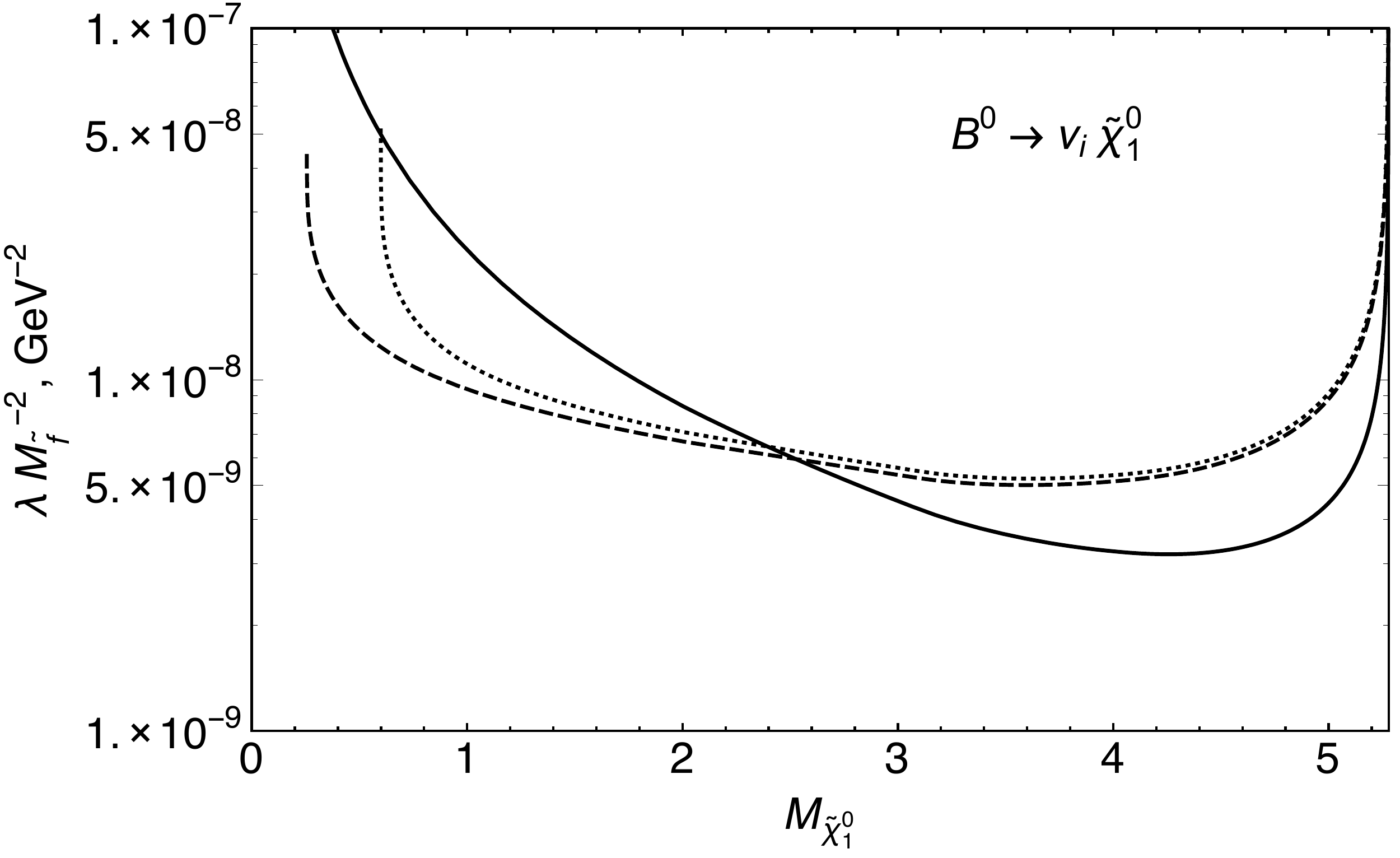}
 \vskip 0.2cm
  \includegraphics[width=0.43\textwidth]{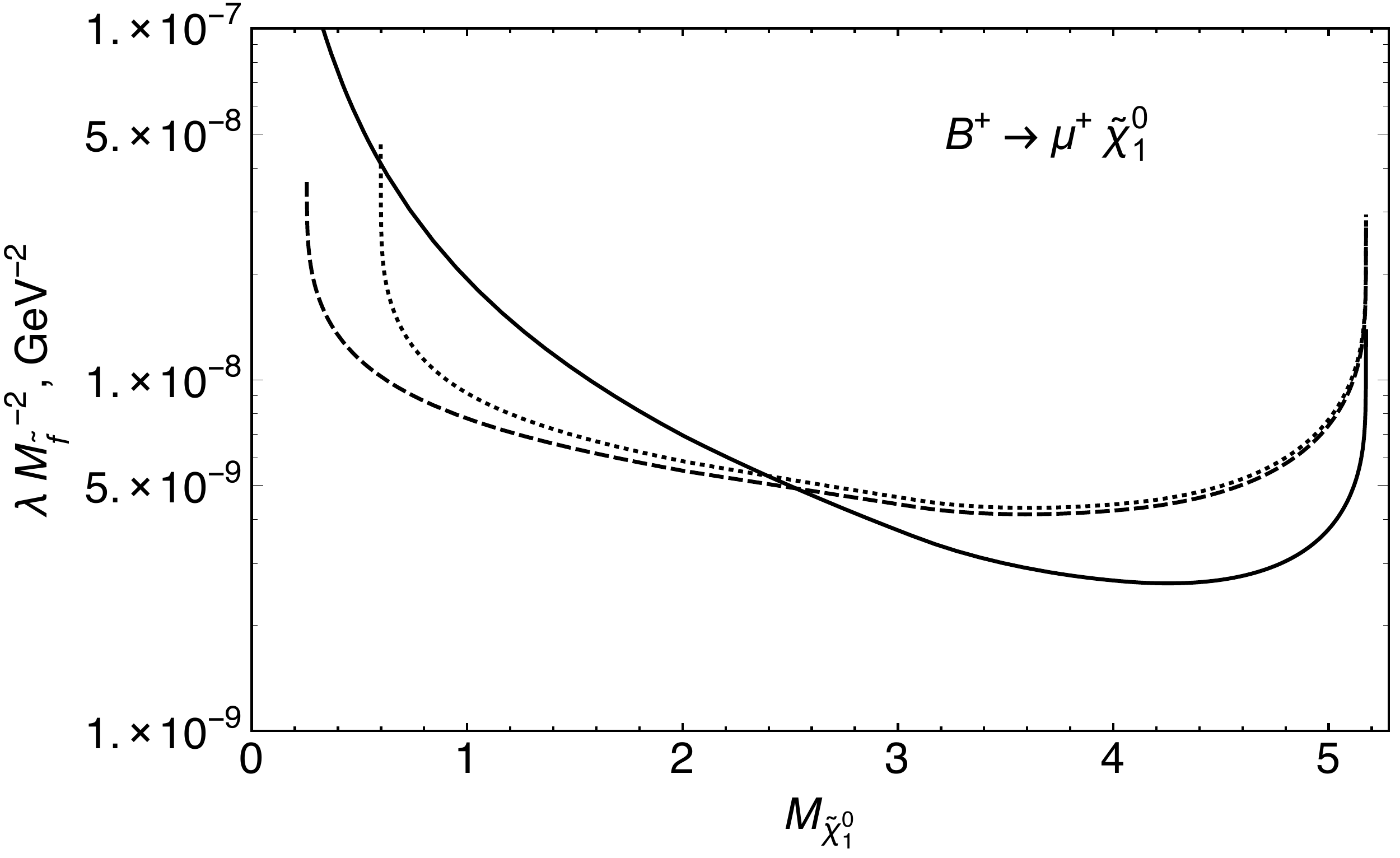}
  \caption{Expected sensitivity of the SHiP experiment to a light
    neutralino with RPV. The solid line refers to leptonic decay, 
the dashed line is for decay to $e^+\pi^-$, and the dotted line is for decay to $e^+ K^-$.
  $\lambda$ stands for the appropriate combination of RPV couplings.}
  \label{for3events}
\end{figure}

Resulting event rates for neutralinos produced in $D^0$ and $B^0$
decays are shown in Fig.\,\ref{LofM}.
Plots for the $D^+$ and $B^+$
channels are very similar to those in Fig.\,\ref{LofM}.  At small
couplings the number of events drops due to poor production, while at
relatively large couplings the number of events also drops because of
fast neutralino decay; thus, one has $l_{\tilde{\chi}_1^0}\ll l_\text{sh}$
and the neutralino flux in the detector is exponentially suppressed 
[see Eq. \eqref{bb}].
The number of events depends both on combination
$(\lambda/M^2_{\tilde{f}})^2$ and on neutralino mass. To demonstrate
this dependence we solve Eq.\,\eqref{Nsig} with $N\equiv 3$ and find
$(\lambda/M^2_{\tilde{f}})^2$ as a function of
$M_{\tilde{\chi}_1^0}$. 
As one can see from Fig.\,\ref{LofM}, there are two solutions to this equations: 
one corresponding to small couplings and slow
decay and another corresponding to relatively large couplings and fast decay. 
Since large couplings are excluded by different
searches (see, e.g., Ref. \cite{Barbier:2004ez} for details), we show in Fig.\,\ref{for3events} a 
solution corresponding to
small couplings.
Note that for the small couplings the probability of neutralino decay inside the fiducial volume \eqref{bb} and, consequently,
the number of events is proportional to the decay width. Therefore, one can use Eqs. \eqref{eqn:B0} -- \eqref{eqn:B+simp} and  Eqs. 
\eqref{leptonic_decay} -- \eqref{semileptonic_decay} (see also expressions from Ref. \cite{Dreiner:1999qz}) in order to
rescale $\lambda'/M^2_{\tilde{f}}$ dependence for the generic case of a nondegenerate mass spectrum and 
any pattern of RPV couplings.

One can see from Fig.\,\ref{for3events} that the decay channel $\tilde{\chi}^0_1\to K^0 \nu_i$
sufficiently affects the lifetime and, subsequently, the event rate 
of neutralinos created in $D$-meson decays via the coupling $\lambda'_{i 2 1}$.

\section{SHiP sensitivity to and CHARM bounds on RPV} 
\label{sec:ship_bounds_on_rpv}

\begin{figure}[!htb]
  \includegraphics[width=0.9\columnwidth]{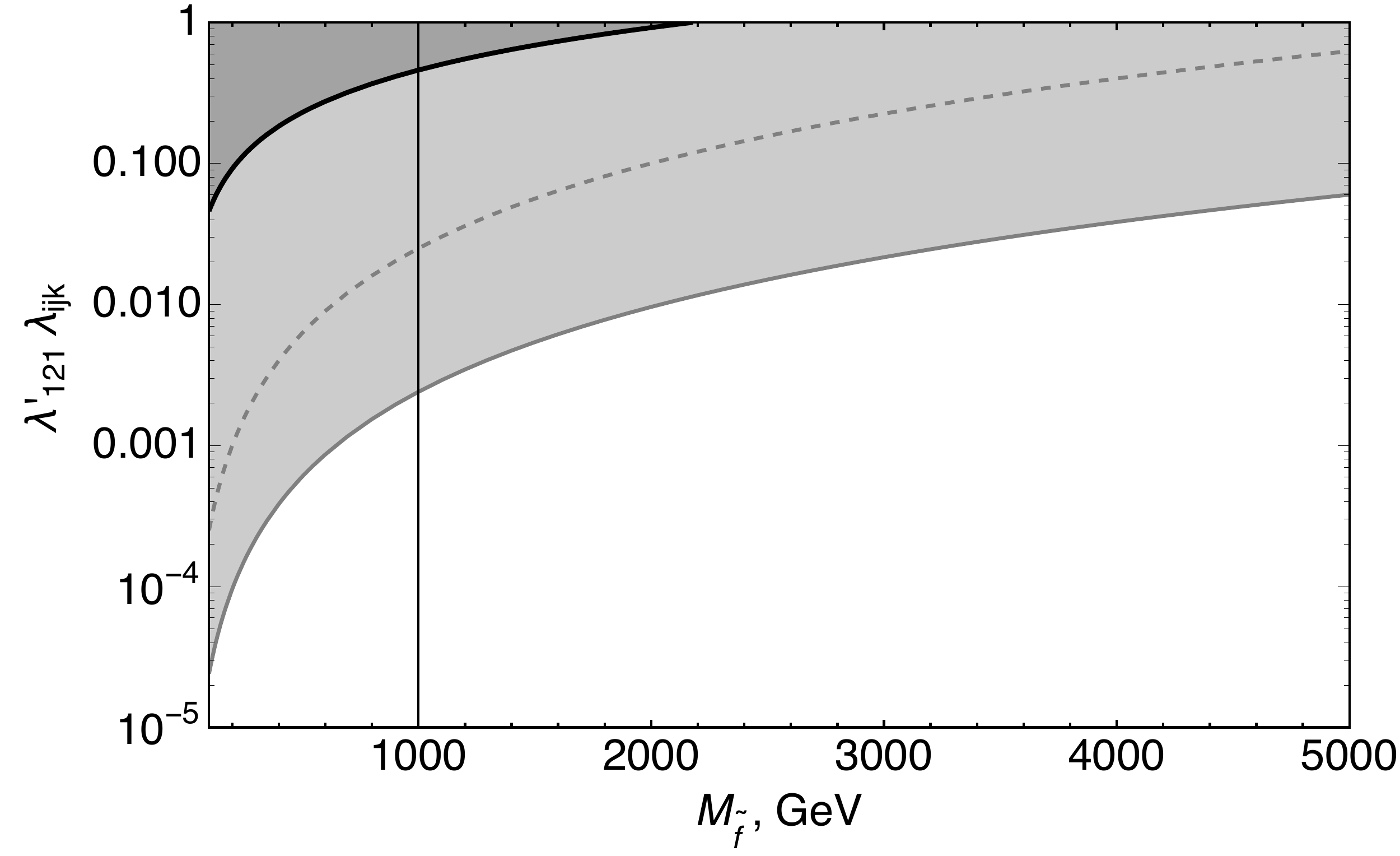}

\vskip 0.4cm
 \includegraphics[width=0.9\columnwidth]{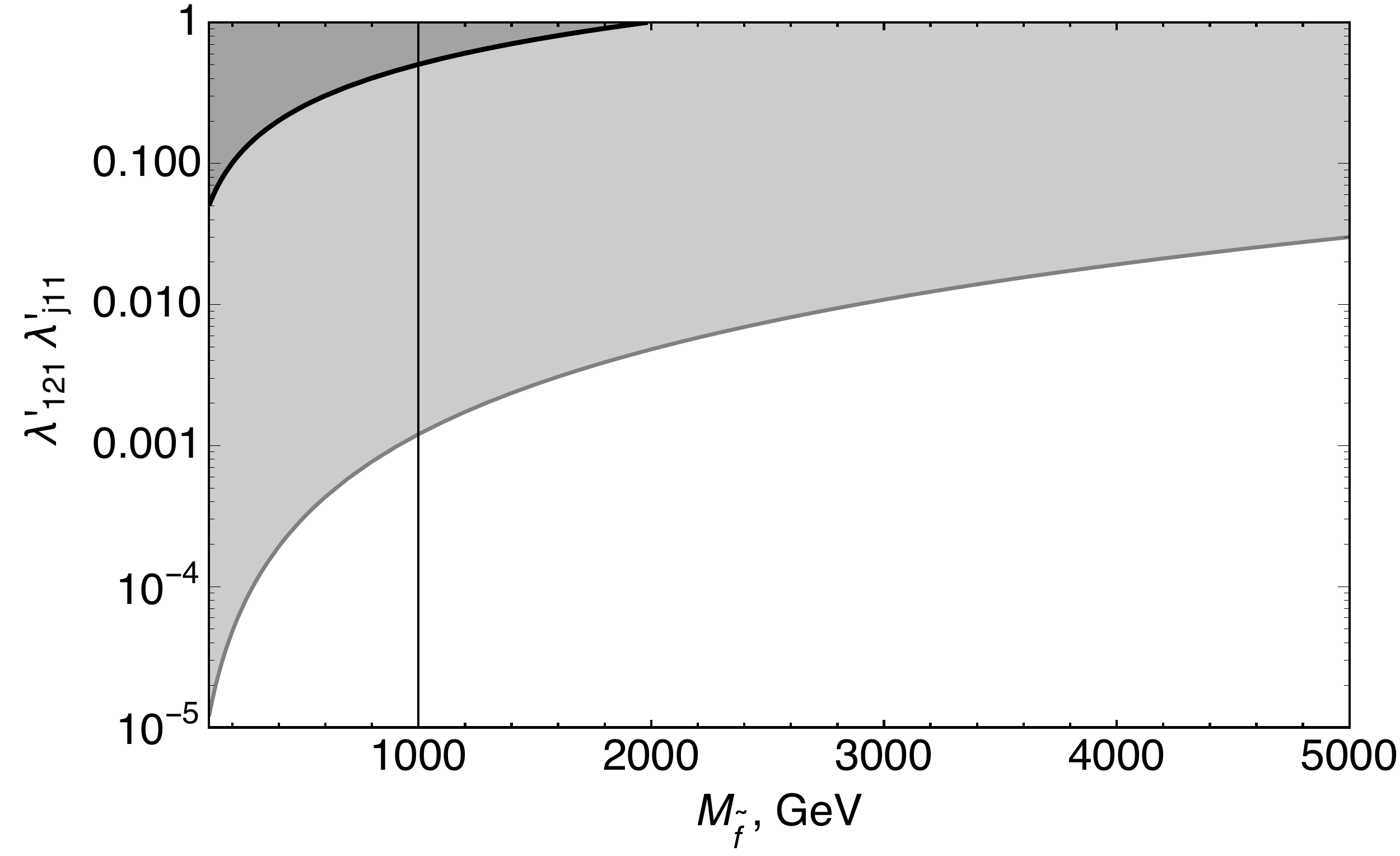}

 \vskip 0.4cm
\includegraphics[width=0.9\columnwidth]{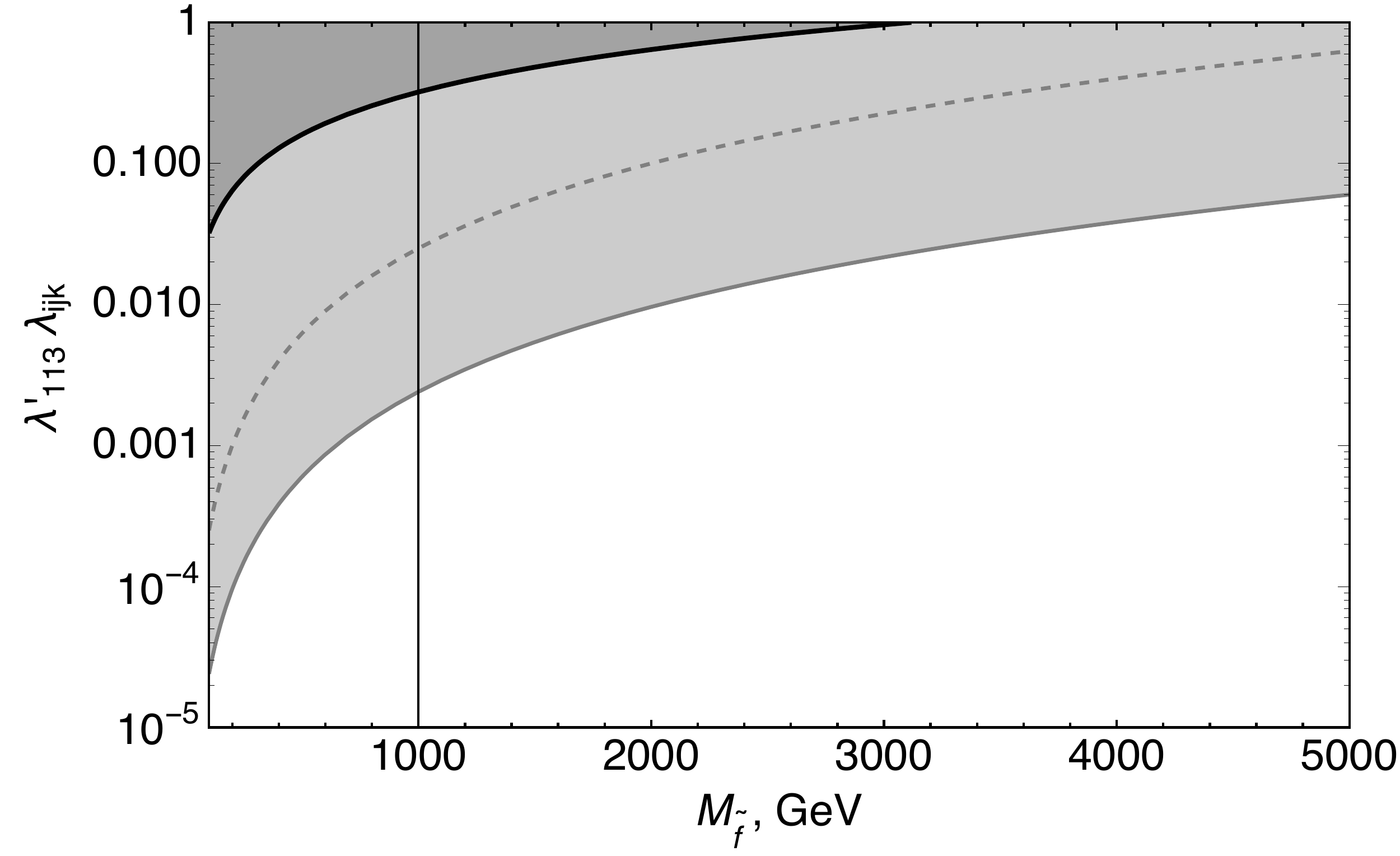}

\vskip 0.4cm
 \includegraphics[width=0.9\columnwidth]{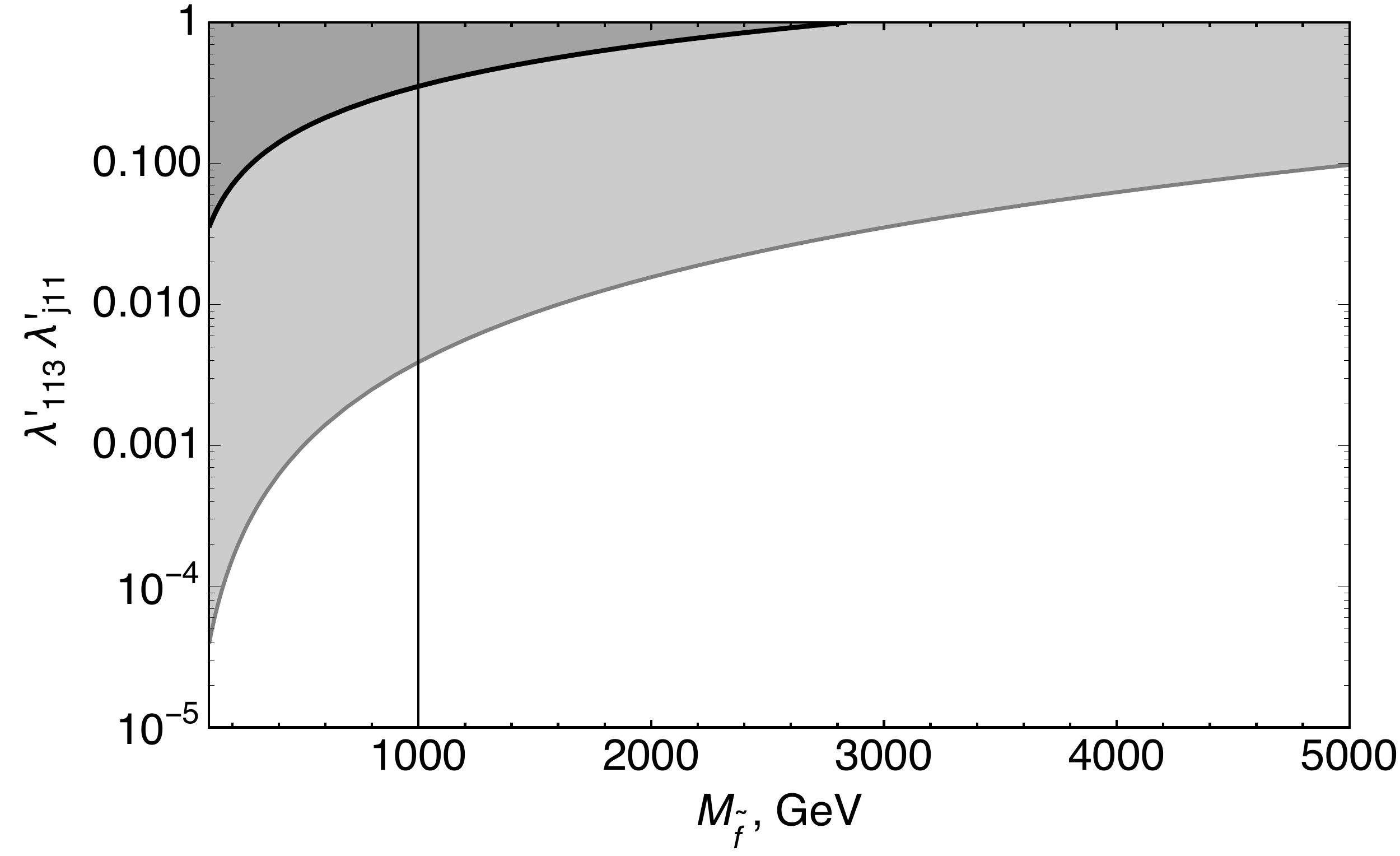}

 \vskip 0.4cm
  \caption{Bounds on $\lambda$ as a function of sfermion mass for the
    light neutralino mass $M_{\tilde{\chi}^0_1}=1$\,GeV. 
The region {\it above} the thick solid black line has been excluded 
by previous studies \cite{Allanach:1999ic}, \cite{Barbier:2004ez}.  
The region {\it above} the gray line could be excluded by SHiP data. The dashed
gray line corresponds to our estimate of CHARM bounds. The region to
the left of the vertical
line at $M_{\tilde{f}}=1$\,TeV is generally disfavored as 
the superpartner mass scale due to searches at the LHC (see discussion in the main text).}
  \label{results}
\end{figure}

Absence of the events, while the three signal events are expected,
implies 95$\%$ confidence level
bounds (if background is negligible, which is true in our
case\,\cite{Anelli:2015pba}).  Limits on RPV couplings that could
be placed by the SHiP experiment depend both on the common sfermion mass
scale and on the neutralino mass.
  Exclusion limits on  various combinations of RPV couplings are shown in
  Fig.\,\ref{results}.  
Dark shaded regions are excluded by previous
studies. Namely, the constraints on $\lambda'_{121}, \lambda'_{113}, \lambda_{121}, \lambda_{122},\lambda_{123}$ have been obtained from charged current universality, and the constraint on
$\lambda'_{111}$ has been obtained from neutrinoless double beta decay (see Refs.
\cite{Allanach:1999ic}, \cite{Barbier:2004ez} for details). Note that 
these constraints scale as $\propto\lambda/M_{\tilde{f}}$, while in
our case of the SHiP sensitivity the scaling goes as 
$\propto\lambda/M_{\tilde{f}}^2$. 

Vertical lines in Fig.\,\ref{results} at $M_{\tilde{f}}=1$\,TeV show the mass scale that is excluded by the LHC searches
in the case of approximately equal sfermion and gluino masses \cite{Asano:2012gj}. Note, however, that for some particular 
RPV spectra this bound can be as low as $800$\,GeV \cite{Graham:2014vya}.

We list our estimates of SHiP bounds on these couplings in Table \ref{table}.
\begin{table}[h]
\caption{Estimates of SHiP sensitivity to and CHARM bounds on combinations of RPV couplings.
In the first three rows we set $M_{\tilde{\chi}_1^0}=1$\,GeV
and $M_{\tilde{\chi}_1^0}=4$\,GeV for the last three rows. Indices $j, k = 1,2$ and $i=1,2,3$ indicate flavor of the final-state leptons.}
\begin{center}
\begin{tabular}{| c | c | c | }

\hline
& Expected sensitivity & Upper limit\\
$\lambda$ & SHiP, $M_{\tilde{f}}^2/\text{TeV}^{2}$ & CHARM, $M_{\tilde{f}}^2/\text{TeV}^{2}$ \\

\hline
$\sqrt{\lambda'_{121}\lambda_{ijk}}$ &  $2.4\times 10^{-3}$ & $2.5\times 10^{-2}$  \\
\hline
$\sqrt{\lambda'_{121}\lambda'_{j11}}$ &  $1.2\times 10^{-3}$ & --  \\
\hline
$\sqrt{\lambda'_{121}\lambda'_{j21}}$ &  $1.4\times 10^{-3}$ &  -- \\
\hline
\hline
$\sqrt{\lambda'_{113}\lambda_{ijk}}$ &  $2.4\times 10^{-3}$ & $2.5\times 10^{-2}$  \\
\hline
$\sqrt{\lambda'_{113}\lambda'_{j11}}$ &  $3.9\times 10^{-3}$ &  -- \\
\hline
$\sqrt{\lambda'_{113}\lambda'_{j21}}$ &  $4.0\times 10^{-3}$ &  -- \\
\hline
\end{tabular}
\end{center}
\label{table}
\end{table}
As one can see from Fig.\,\ref{for3events} the bounds listed in Table
\ref{table} are valid for a wide range of kinematically allowed region
(with phase-space corrections at the boundaries) of $M_{\tilde{\chi}_1^0}$.  

In order to illustrate advantages of the SHiP facility, we also present in
Fig.\,\ref{results} our estimates of the bounds on RPV couplings that
follow from the absence of signal in the CHARM experiment.  These
bounds are obtained by adapting the whole procedure described above
for the CHARM geometry \cite{Bergsma:1983rt}.  The CHARM experiment
has exploited the same $400$ GeV beam as the SHiP plans.  The detector
was located at 480\,m downstream from the beam dump.  Therefore, it covers a
sufficiently smaller solid angle compared to the SHiP. The length of
the decay region was 35\,m and the radius of the calorimeter placed at the end of
decay volume was 1.5\,m.  The total amount of protons on target equalled
$2.4 \times 10^{18}$ \cite{Bergsma:1983rt}.  To the best of our
knowledge there was no special investigation of the CHARM sensitivity to
RPV SUSY, but very similar signatures of heavy neutral lepton decays to the SM leptons were
studied in Refs. \cite{Bergsma:1983rt}, \cite{Bergsma:1985is}.

\section{Conclusion}
\label{sec:Conclusion}
To summarize, we have estimated the sensitivity of the recently proposed SHiP
experiment to the supersymmetric extensions of the SM with light
neutralinos and $R$-parity violation. 
For the $R$-parity violating
couplings $\lambda$ of order one, the SHiP will allow us to probe the superpartner
mass scale as high as 30\,TeV (see Fig.\,\ref{for3events}), which is in
agreement with previous estimates in Ref.\,\cite{Alekhin:2015byh}.
The number of signal events scales as $ \propto (\lambda'/M^2_{\tilde{f}})^4$.
As a by-product we have obtained limits on the model parameters from
nonobservation of anomalous events in the CHARM experiment. With
respect to the CHARM, the SHiP will
improve the sensitivity to $R$-parity-violating couplings by an order
of magnitude.  

Several remarks are in order. First, other final states such as the 
mentioned neutral kaons must be considered as well. Second, the light
neutralinos can be produced in pairs due to $R$-parity-conserving
couplings, and the corresponding new production channels can also be
studied. Third, not only heavy meson but also heavy baryons can
decay into light neutralinos, which gives additional production
channels. Fourth, secondary hadrons, produced in the
hadron showers initiated by 400\,GeV proton scattering off target
materials, can contribute to the light neutralino production. Fifth,
$\tau$ leptons produced mostly in decays of $D_s$ mesons allow us to
probe other $R$-parity-violating couplings $\lambda_{ijk}$, which is
also worth investigating.

This work
has been supported by Russian Science Foundation Grant No.
14-22-00161.
\bibliography{refs}

\end{document}